\documentclass[
    ,final            
  ]
  {aipproc}

\layoutstyle{6x9}

\def\simlt{\mathrel{\hbox{\rlap{\hbox{\lower4pt\hbox{$\sim$}}}\hbox{$<$}}}}
\def\simgt{\mathrel{\hbox{\rlap{\hbox{\lower4pt\hbox{$\sim$}}}\hbox{$>$}}}}

\begin{document}

\title{The Afterglows and Host Galaxies of Short GRBs: An Overview}


\author{Edo Berger}{
  address={Carnegie Observatories, Pasadena, CA}}

\begin{abstract}
Despite a rich diversity in observational properties, gamma-ray bursts
(GRBs) can be divided into two broad categories based on their
duration and spectral hardness -- the long-soft and the short-hard
GRBs.  The discovery of afterglows from long GRBs in 1997, and their
localization to arcsecond accuracy, was a watershed event.  The
ensuing decade of intense study led to the realization that long-soft
GRBs are located in star forming galaxies, produce about $10^{51}$ erg
in collimated relativistic ejecta, are accompanied by supernovae, and
result from the death of massive stars.  While theoretical arguments
suggest that short GRBs have a different physical origin, the lack of
detectable afterglows prevented definitive conclusions.  The situation
changed dramatically starting in May 2005 with the discovery of the
first afterglows from short GRBs localized by \emph{Swift} and 
\emph{HETE-2}.  Here I summarize the discovery of these afterglows 
and the underlying host galaxies, and draw initial conclusions about
the nature of the progenitors and the properties of the bursts.
\end{abstract}

\maketitle


\section{History and Models}

The detection of short-duration gamma-ray bursts (GRBs) dates back to
the \emph{Vela} satellites \cite{strong:74}.  However, only in 1993
the short bursts (with $T_{90}\simlt 2$ s) were recognized as a
separate sub-class from the long GRBs, and were furthermore shown to
have on average a harder $\gamma$-ray spectrum \cite{kouveliotou:93};
hereafter I will refer to short-hard bursts as SHBs.  The short
durations of SHBs suggest that they are unlikely to result from the
death of massive stars, for which the natural timescale (the free-fall
time) is significantly longer, $t_{ff}\approx 30\,{\rm
s}\,(M/10\,M_\odot)^{-1/2}(R/10^{10}\,{\rm cm})^{3/2}$.

Instead the main theoretical thrust has been focused on coalescing
compact objects -- neutron stars and/or black holes (DNS or NS-BH) --
as the progenitors of SHBs
\cite{eichler:89,paczynski:91,narayan:92,katz:96,rosswog:03}.  In this
context the duration, which is set by the viscous timescale of the gas
accreting onto the newly-formed black hole, is short due to the small
scale of the system.  Other progenitors have been proposed in addition
to the DNS and NS-BH systems, namely magnetars, thought to be the
power source behind soft $\gamma$-ray repeaters \cite{thompson:95},
and accretion-induced collapse (AIC) of neutron stars
\cite{qin:98,macfadyen:05}.  The magnetar model is unlikely to account
for SHBs at cosmological distances, since even the 2004 Dec.~27 giant
flare from SGR\,1806-20 would only be detected by \emph{BATSE} or 
\emph{Swift} at $\simlt 50$ Mpc; magnetars may contribute to a local 
population of SHBs.  The AIC model has not been investigated in
detail, but in the case of a white dwarf was shown to be too baryon
rich to produce GRBs \cite{fryer:99}.

Naturally, without a distance and energy scale, or an understanding of
the micro- and macro-environments of SHBs, it is nearly impossible to
make any quantitative statements about their progenitors or the
detailed underlying physics.  As in the case of the long GRBs, this
understanding relies on arcsecond positions, which in turn require the
identification of afterglows.  Several SHBs have been localized to
sufficient accuracy (few arcmin$^2$) in the past to allow afterglow
searches, but none have been detected at optical or radio wavelengths
due to delayed and shallow searches \cite{hurley:02}.

\section{The Discovery of Short GRB Afterglows}

The breakthrough in understanding the origin and properties of SHBs
resulted from the localization of the first afterglows starting in May
2005.  Here I provide a short account of these discoveries,
highlighting the growing understanding achieved with each burst.

\begin{figure}
\includegraphics[height=.45\textheight,angle=270]{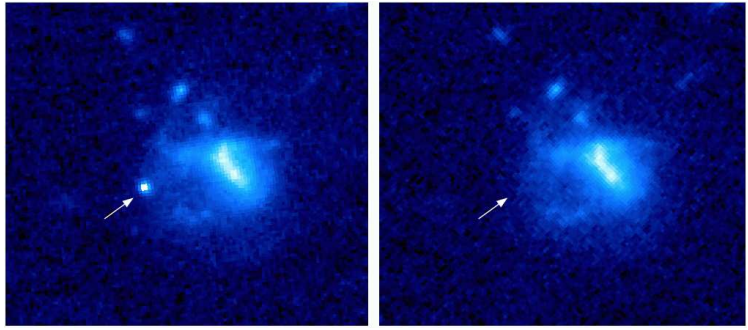}
\caption{HST images of the optical afterglow and irregular star 
forming host of GRB\,050709.}
\label{fig:050709}
\end{figure}

\medskip

\noindent{\bf GRB\,050509b} was detected by \emph{Swift} on 2005 May 
9.167 UT with a duration of $40$ ms and a fluence, $F_\gamma=9.5\times
10^{-9}$ erg cm$^{-2}$ \cite{gehrels:05}.  A fading X-ray afterglow
was localized with the XRT onboard \emph{Swift} to a positional
accuracy of $9.3''$ radius \cite{gehrels:05}.  Despite intense
follow-up at optical and radio wavelengths no candidates were
discovered to a limit of 24 mag at $t\approx 2.5$ hr
\cite{bloom:05,hjorth:05a} and 0.1 mJy at $t\approx 2.2$ hr
\cite{gcn:3387}, respectively.  Instead, a bright elliptical 
galaxy ($L\approx 3\,L^*$) at $z=0.225$ was detected in coincidence
with the X-ray error circle \cite{bloom:05,gehrels:05}.  The \emph{a
posteriori} probability of such a coincidence has been estimated at
$0.01-1\%$ \cite{bloom:05,gehrels:05,pedersen:05}.  The implication of
this possible association is that the progenitors of SHBs are related
to an old stellar population.  Unfortunately, the XRT error circle
also contains over twenty fainter galaxies, which most likely reside
at higher redshifts, thereby preventing a definitive association.  The
optical follow-up also placed a limit of $M_B>-13.3$ mag (5 mag
fainter than SN\,1998bw) on a supernova coincident with GRB\,050509b
\cite{bloom:05,hjorth:05a}, providing additional support to a
non-massive star origin \emph{if the redshift of $z=0.225$ is
adopted}.

\begin{figure}
\includegraphics[height=.19\textheight]{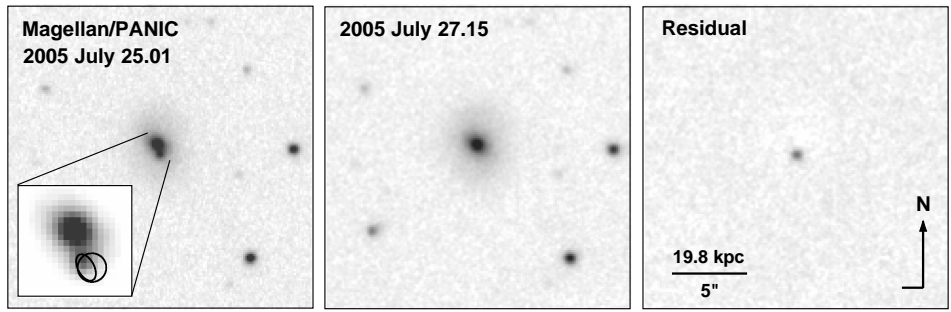}
\caption{Magellan near-IR images of the afterglow and host galaxy 
of GRB\,050724 on two separate occasions, and the residual image
clearly showing the fading afterglow.  The inset shows the radio
(ellipse) and X-ray (circle) positions of the afterglow.  From
Ref.~\cite{berger:05}.}
\label{fig:050724}
\end{figure}

\medskip

\noindent{\bf GRB\,050709} was detected by \emph{HETE-2} on May 
9.942 UT with a duration of $70$ ms and a fluence, $4.0\times 10^{-7}$
erg cm$^{-2}$ ($2-400$ keV).  The initial pulse was followed $25$ s
later by a softer X-ray bump with a duration of $130$ s and a fluence
of $1.1\times 10^{-6}$ erg cm$^{-2}$ ($2-25$ keV)
\cite{villasenor:05}.  Observations with \emph{Chandra} provided an
arcsecond position \cite{fox:05}, and led to the discovery of the
optical afterglow \cite{hjorth:05b,fox:05} (Fig.~\ref{fig:050709}).
No radio afterglow was detected \cite{fox:05}.  Spectroscopy revealed
that the host is a star forming galaxy at $z=0.160$, but HST imaging
showed that the burst did not coincide with a bright star forming
region \cite{fox:05}.  No supernova component was detected
\cite{hjorth:05b}.  These results suggest that while GRB\,050709
occurred in a star forming galaxy, it was not related to a massive
star.

\medskip

\noindent{\bf GRB\,050724} was localized by \emph{Swift} on 2005 
July 24.524 UT with a duration of $3\pm 1$ s, dominated by an initial
spike of 250 ms duration, and a fluence, $F_\gamma=3.9\times 10^{-7}$
erg cm$^{-2}$ ($15-150$ keV) \cite{barthelmy:05}.  As in the case of
GRB\,050709, the initial pulse was followed 30 s later by a soft bump
($15-25$ keV) which lasted 120 s, but with a fluence of only $10\%$
that of the main pulse.  A bright X-ray afterglow detected by XRT
\cite{barthelmy:05} led to the discovery of the radio, optical
and near-IR afterglow, which coincided with a bright elliptical galaxy
at $z=0.257$ (Fig.~\ref{fig:050724}) \cite{berger:05}.  This burst
provided the first unambiguous association with a galaxy undergoing no
current star formation; the limit at the position of the GRB is
$<0.05$ M$_\odot$ yr$^{-1}$ \cite{berger:05}.  The X-ray and optical
light curves revealed a broad increase by nearly an order of magnitude
at 0.65 d, suggestive of energy injection
\cite{barthelmy:05,berger:05}.  The subsequent steep decline at 
$t\simgt 1$ d is reminiscent of the jet breaks detected in long GRBs,
indicating a jet opening angle of about $8.5^\circ$ \cite{berger:05}.

\medskip

\noindent{\bf GRB\,050813} was detected by \emph{Swift} on 2005 August 
13.281 UT with a duration of 0.6 s and a fluence, $4.3\times 10^{-8}$
erg cm$^{-2}$ ($15-150$ keV).  The initial X-ray position included a
pair of galaxies, of which one is at a redshift $z=0.72$
\cite{gcn:3801,gcn:3808}, and appears to be part of a galaxy cluster 
\cite{gcn:3798}.  However, the revised XRT position excludes these 
galaxies, and instead contains a galaxy at $z\approx 1.8$, which
appears to be part of a cluster at that redshift.  No optical or radio
afterglows were detected.  Since this is the highest redshift SHB to
date, it has significant implications for the progenitor lifetime (see
below).

\begin{figure}
\includegraphics[height=.6\textheight,angle=270]{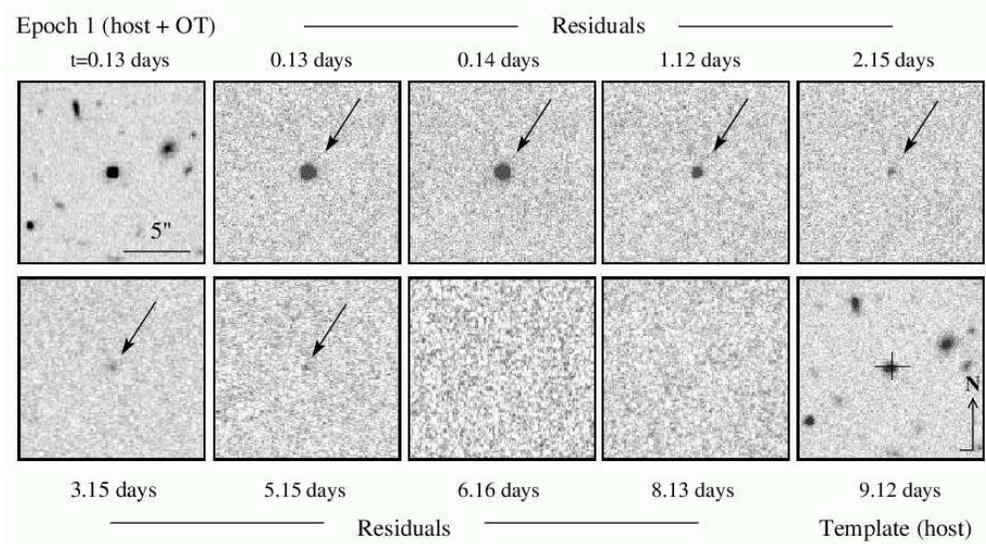}
\caption{The fading optical afterglow and host galaxy of GRB\,051221.
From Ref.~\cite{soderberg:06}.}
\label{fig:051221}
\end{figure}

\medskip

\noindent{\bf GRB\,051221} is perhaps the best-studied SHB to date.  
It was discovered by \emph{Swift} on 2005 December 21.077 with an
initial hard pulse of 250 ms duration, followed by softer emission,
which lasted about 1 s, and a total fluence, $3.2\times 10^{-6}$ erg
cm$^{-2}$ ($20-2000$ keV) \cite{gcn:4363,gcn:4394}.  The X-ray
afterglow was localized to $3.5''$ radius accuracy \cite{gcn:4366},
leading to the identification of the optical, near-IR, and radio
counterparts \cite{soderberg:06} (Fig.~\ref{fig:051221}).  A spectrum
of the combined afterglow and host indicated a redshift of $z=0.546$
\cite{soderberg:06}.  The afterglow evolution follows a simple power
law decay to $t\approx 13$ d, interrupted only by a period of
flattening from 1.4 to 3.1 hr in the X-rays and reverse shock emission
in the radio band (Fig.~\ref{fig:051221-lcs}).  The light curve
evolution indicates an opening angle $>13^\circ$, an energy injection
of about a factor of three, and a total energy of at least $1.0\times
10^{50}$ erg \cite{soderberg:06}.

\section{The Properties and Progenitors of Short GRBs}

The discovery of SHB afterglows allows us to address for the first
time the basic questions regarding the nature and properties of these
events:
\begin{itemize}
\item What is the energy release of SHBs (prompt emission and 
relativistic blast wave)?
\item Is there evidence for prolonged engine activity?
\item Are the ejecta collimated or spherical?
\item Are SHBs accompanied by supernova-like events?
\item What is the density and structure of the circumburst medium?
\item In what type of host galaxies do SHBs occur?
\end{itemize}

\subsection{Short GRB Energetics}

\begin{figure}
\includegraphics[height=.25\textheight]{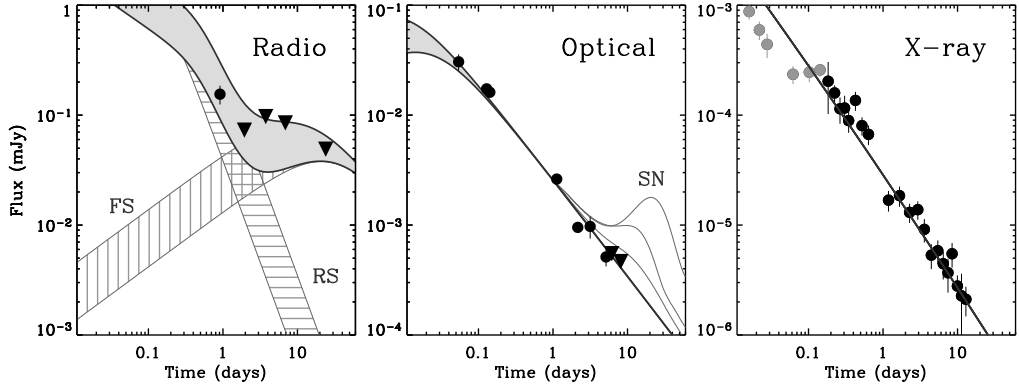}
\caption{Broad-band light curves of the afterglow of GRB\,051221.  
The black lines and shaded regions are model fits, which include
forward shock emission and the reverse shock contribution in the
radio, which is associated with the energy injection episode in the
X-rays (at 0.05 d).  From Ref.~\cite{soderberg:06}.}
\label{fig:051221-lcs}
\end{figure}

\begin{figure}
\resizebox{18.3pc}{!}{\includegraphics[height=.5\textheight,angle=90]{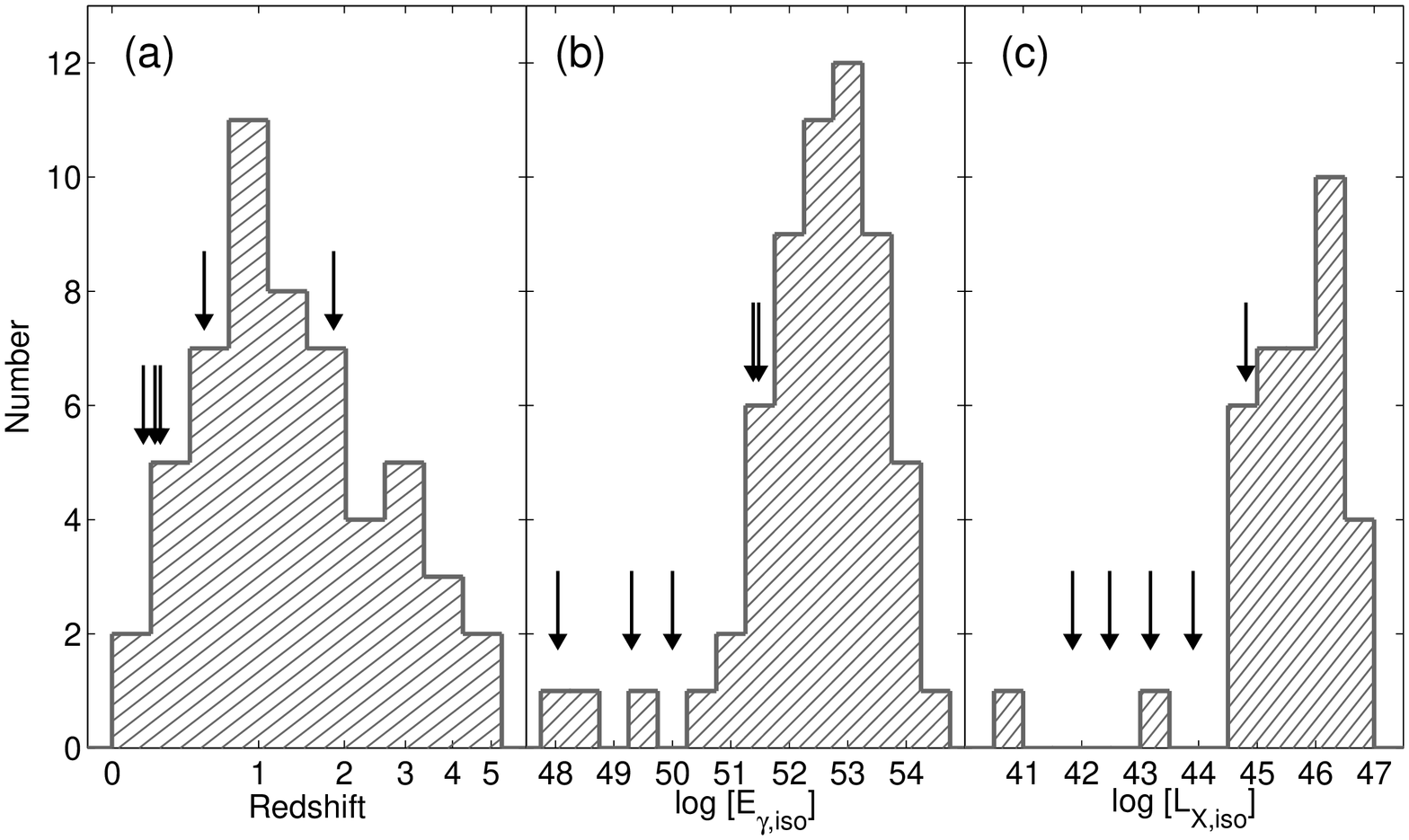}}
\resizebox{13.9pc}{!}{\includegraphics[height=.5\textheight]{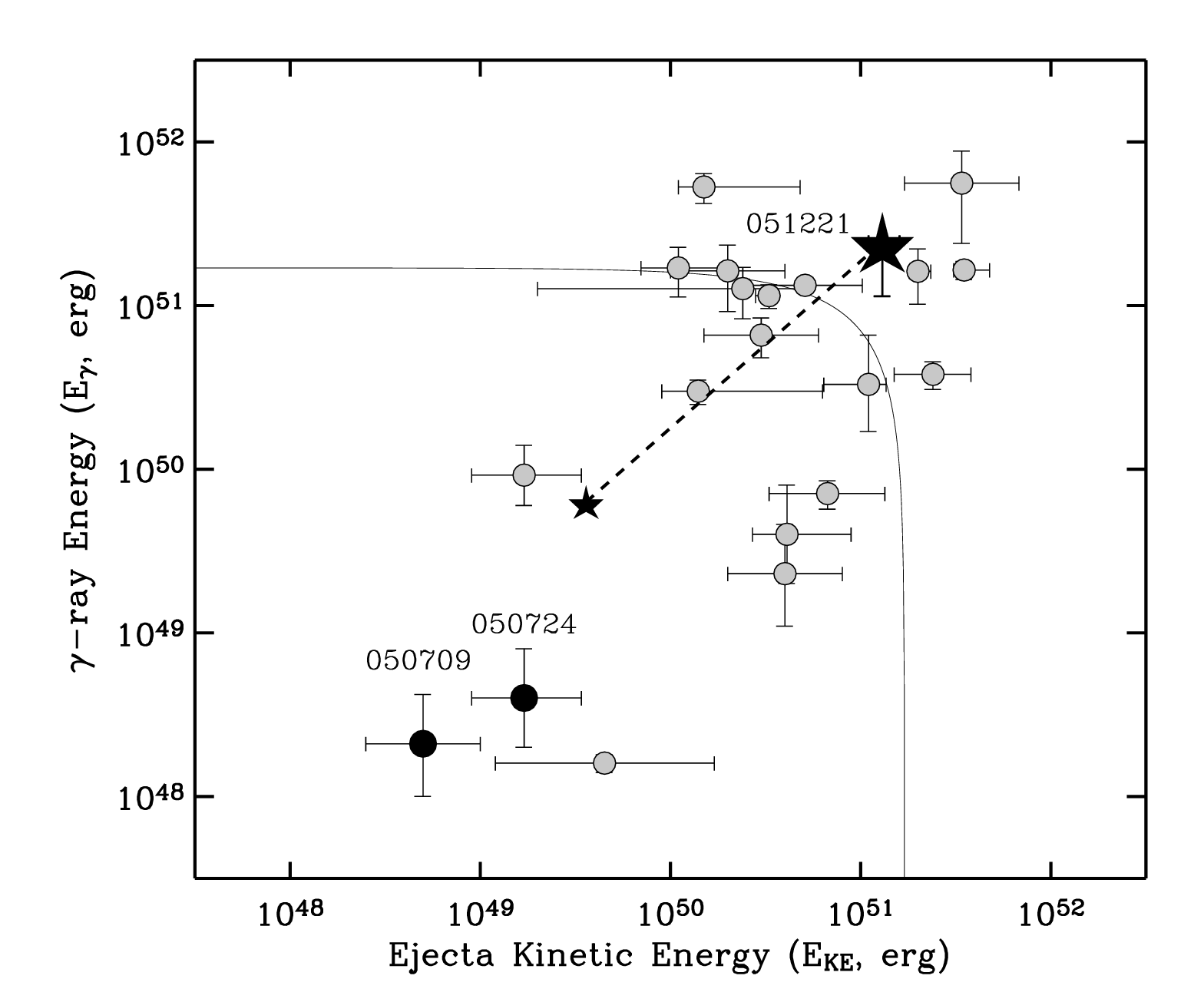}}
\caption{\emph{Left:} Distributions of (a) redshift, (b) isotropic 
$\gamma$-ray energy, and (c) isotropic X-ray luminosity (a proxy for
the blast wave kinetic energy).  Adapted from Ref.~\cite{fox:05}.
\emph{Right:} Beaming corrected $\gamma$-ray and afterglow kinetic
energies for SHBs (black) and long GRBs (gray).  From
Ref.~\cite{soderberg:06}.}
\label{fig:energy}
\end{figure}

The redshift distribution of the five SHBs with precise or putative
redshifts ranges from about 0.16 to $\sim 1.8$, with three of the five
bursts at $z\simlt 0.3$ (Fig.~\ref{fig:energy}); GRB\,050813 stands
out with a significantly higher redshift.  At the same time, a
statistical comparison of BATSE SHB positions to low-redshift galaxy
catalogs suggests that $\sim 10-25\%$ of the BATSE SHBs may originate
within 100 Mpc \cite{tanvir:05}.  The isotropic $\gamma$-ray energies
span a wide range, $E_{\rm \gamma,iso}\approx 1.1\times 10^{48} -
3\times 10^{51}$ erg, which is at the low end of the distribution for
long GRBs (Fig.~\ref{fig:energy}) \cite{fox:05}.  The afterglow X-ray
luminosities, which serve as a proxy for the blast-wave kinetic energy
\cite{freedman:01}, span an equally wide range, $L_{X,{\rm iso}}
(t=10\,{\rm hr})\approx 7.0\times 10^{41}-6.4\times 10^{44}$ erg
s$^{-1}$.  These values are again at the low end of the distribution
for long GRBs \cite{berger:03}.

As in the case of long GRBs, the true energy release is strongly
dependent on collimation of the ejecta.  Both GRBs 050709 and 050724
exhibit evidence for jets through breaks and steep decays of their
afterglow light curves \cite{berger:05,fox:05}.  In the former case
the opening angle is $14^\circ$ while for the latter it is about
$8.5^\circ$.  GRB\,051221, on the other hand, exhibits no clear break
out to at least 13 days, or $\theta_j>13^\circ$ \cite{soderberg:06}.
The inferred jet angles are consistently wider than the median value
of $\langle\theta_j\rangle\approx 5^\circ$ for long GRBs
\cite{soderberg:06}.

The broad-band light curves of GRBs 050709, 050724, and 051221 also
allow a rough determination of the blast wave kinetic energies using
the standard synchrotron model \cite{sari:98,soderberg:06}
(Fig.~\ref{fig:051221-lcs}).  The parameters of interest are $E_{\rm
KE}$, the circumburst density ($n$), and the fractions of energy in
relativistic electron ($\epsilon_e$) and magnetic fields
($\epsilon_B$).  The values derived for the three SHBs are summarized
in Tab.~\ref{tab:model}, and the derived beaming corrected energies
are plotted in Fig.~\ref{fig:energy}
\cite{berger:05,fox:05,soderberg:06}.  The energies of GRBs 050709 and
050724 are about two orders of magnitude lower than those of long
GRBs, $E\approx {\rm few}\times 10^{51}$ erg
\cite{frail:01,bloom:03,berger:03b}, but GRB\,051221 has an energy 
that is at least an order of magnitude larger, and potentially as
large as that of long GRBs.

The energy scale of GRB\,051221 has important implications for the
energy extraction mechanism, $\nu\bar{\nu}$ annihilation or MHD
processes related to the black hole and/or the accretion disk
\cite{lee:02,rosswog:03}.  Numerical calculations reveal that
$\nu\bar{\nu}$ annihilation is unlikely to power an SHB with a
beaming-corrected energy in excess of ${\rm few}\times 10^{48}$ erg,
while MHD processes can produce $>10^{52}$ erg s$^{-1}$.  Thus, GRBs
050709 and 050724 could in principle be powered by $\nu\bar{\nu}$
annihilation, but it is unlikely that GRB\,051221 was
\cite{soderberg:06}.

\subsection{Prolonged Engine Activity}

\begin{figure}
\resizebox{17pc}{!}{\includegraphics[height=.5\textheight]{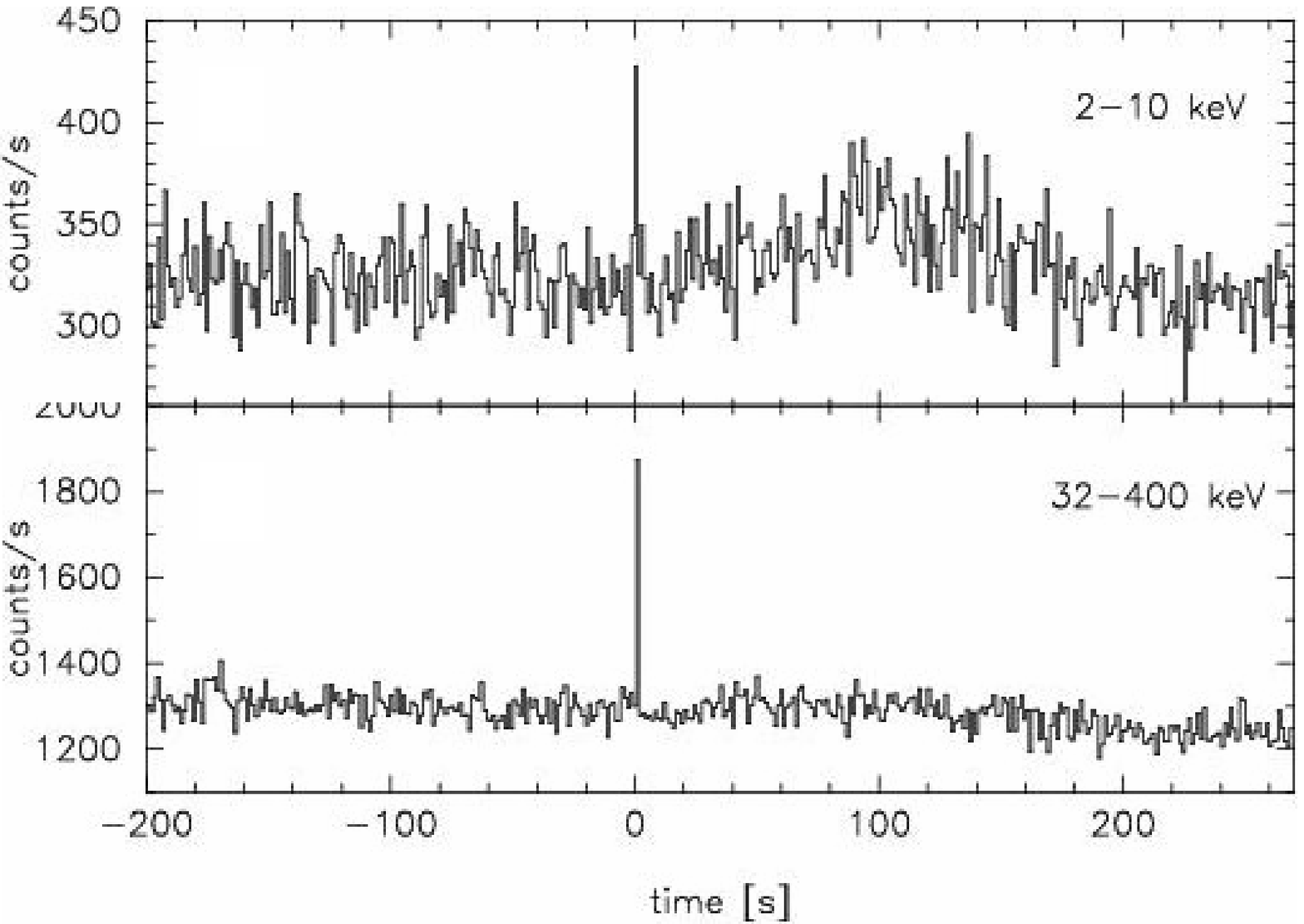}}
\resizebox{16pc}{!}{\includegraphics[height=.5\textheight]{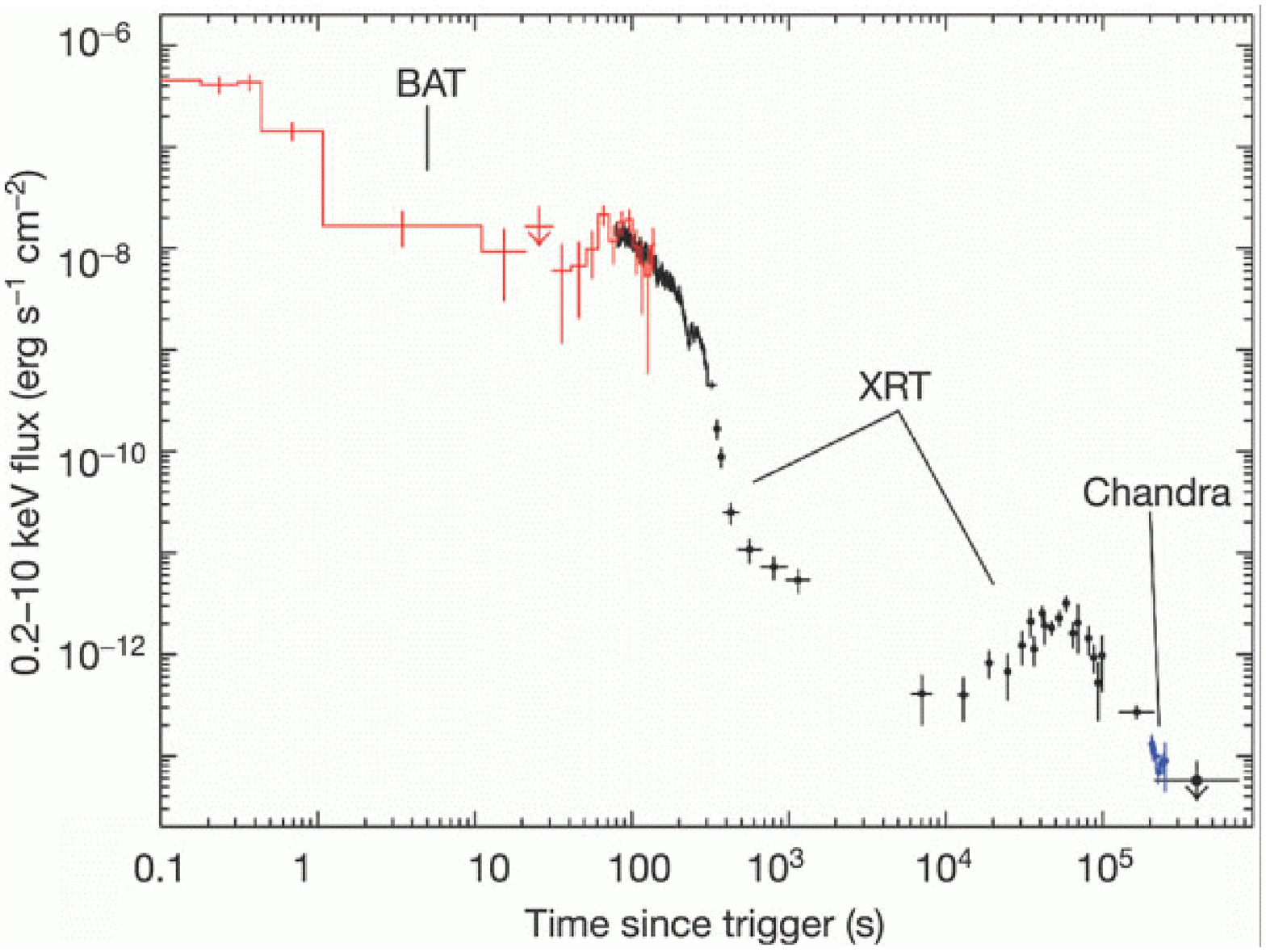}}
\caption{Flares in the prompt emission and X-ray afterglow of GRBs 
050709 (left; Ref.~\cite{villasenor:05}) and 050724 (right;
Ref.~\cite{barthelmy:05}).}
\label{fig:flares}
\end{figure}

One of the intriguing results emerging from observations of the prompt
emission and X-ray afterglows is evidence for prolonged engine
activity and delayed energy injection.  A hint of this result was
already available from summed light curves of BATSE SHBs, which led to
the possible detection of excess emission peaking $\sim 30$ s after
the burst with a duration of $\sim 100$ s
\cite{lazatti:01,connaughton:02}.  This was originally proposed as 
evidence for afterglow emission, but there is now evidence from GRBs
050709 and 050724 that this is not the case (Fig.~\ref{fig:flares}).
In addition to the early flares, the light curves of GRBs 050724 and
051221 exhibit evidence for delayed energy injection
\cite{barthelmy:05,soderberg:06}, which may arise from extended engine
activity and/or a wide distribution of ejecta Lorentz factors.

\begin{table}
\begin{tabular}{lccc}
\hline
\tablehead{1}{r}{b}{}
& \tablehead{1}{c}{b}{050709}
& \tablehead{1}{c}{b}{050724}
& \tablehead{1}{c}{b}{051221} 
\\\hline
Redshift                   & $0.160$             & $0.257$              & $0.546$ \\
$E_{\rm \gamma,iso}$ (erg) & $6.9\times 10^{49}$ & $4.0\times 10^{50}$  & $2.4\times 10^{51}$ \\
$E_{\rm KE,iso}$ (erg)     & $1.6\times 10^{48}$ & $1.5\times 10^{51}$  & $1.4\times 10^{51}$ \\
$n$ (cm$^{-3}$)            & $\sim 0.01$   	 & $\sim 0.1$           & $\sim 10^{-3}$ \\
$\epsilon_e$	           & $\sim 0.3$	         & $\sim 0.04$          & $\sim 0.3$ \\
$\epsilon_B$	           & $\sim 0.3$	         & $\sim 0.02$          & $\sim 0.1$ \\
$\theta_j$ (deg)	   & $14$	         & $9$                  & $>13$ \\
$f_b$		           & $0.03$	 	 & $0.01$               & $>0.03$ \\
$E_\gamma$ (erg)	   & $2.1\times 10^{48}$ & $4.5\times 10^{48}$  & $(6-240)\times 10^{49}$ \\
$E_{\rm KE}$ (erg)	   & $5.0\times 10^{48}$ & $1.7\times 10^{49}$  & $(4-140)\times 10^{49}$ \\
Reference	           & \cite{fox:05}       & \cite{berger:05}     & \cite{soderberg:06} \\\hline
\end{tabular}
\caption{Physical Properties of SHBs}
\label{tab:model}
\end{table}

Given the short lifetime and dynamical timescale in the case of DNS or
NS-BH mergers, the theoretical expectation is for simple afterglow
evolution and a short duration of the $\gamma$-ray emission.  The
X-ray flares do not fit naturally in this scenario, although delayed
activity arising from fragmentation of the accretion disk has been
proposed and argued to agree with the basic observational signatures
\cite{perna:06}.  An intriguing alternative, in the context of
accretion-induced collapse, is that the flares result from the
interaction of the emerging blast wave with the giant companion star,
with the timescale and duration of the flare set by the binary
separation and the radius of the companion \cite{macfadyen:05}.  Two
observations can be used to test this scenario: (i) multiple flares,
and (ii) if most SHBs are collimated, then only the small fraction
whose jets intercept the companion should exhibit flares.

\subsection{Host Galaxies and Offsets}

One of the main clues that long GRBs are related to the death of
massive stars came from their association with star forming galaxies.
Long GRB hosts have star formation rates (SFR) typically in excess of
1 M$_\odot$ yr$^{-1}$ (and sometimes $>100$ M$_\odot$ yr$^{-1}$
\cite{berger:03c}), stellar populations younger than $\sim 10^8$ yr, 
and specific star formation rates that are higher than the general
population of star forming galaxies \cite{christensen:04}.  In
addition, the distribution of long GRBs relative to their hosts traces
that of massive stars \cite{bloom:02}.

In a similar vein, we can use the hosts of SHBs to address the
properties of the progenitors.  The most striking difference compared
to long GRBs is the existence of SHBs in elliptical galaxies.  The
limits on the SFR for the elliptical hosts of GRBs 050509b and 050724
are $<0.1$ and $<0.05$ M$_\odot$ yr$^{-1}$, respectively
\cite{bloom:05,berger:05}.  Even the star forming hosts of GRBs
050709 and 051221, with $\sim 0.5$ and $\sim 1.5$ M$_\odot$ yr$^{-1}$,
respectively, have lower SFRs than the median for hosts of long GRBs
\cite{fox:05,soderberg:06}.  The host of GRB\,051221, moreover, 
exhibits evidence for an evolved stellar population and a near solar
metallicity.

The mix of host types is reminiscent of type Ia supernovae (SNe Ia),
which occur in both early- and late-type galaxies.  It is interesting
to note that the luminosity of SNe Ia is correlated with the host
type, with intrinsically dimmer events occurring in early-type
galaxies \cite{hamuy:00}.  The current sample of SHBs is too small to
address possible correlations of the burst and host properties, but
this should become possible in the future.

\begin{figure}
\resizebox{18pc}{!}{\includegraphics[height=.5\textheight]{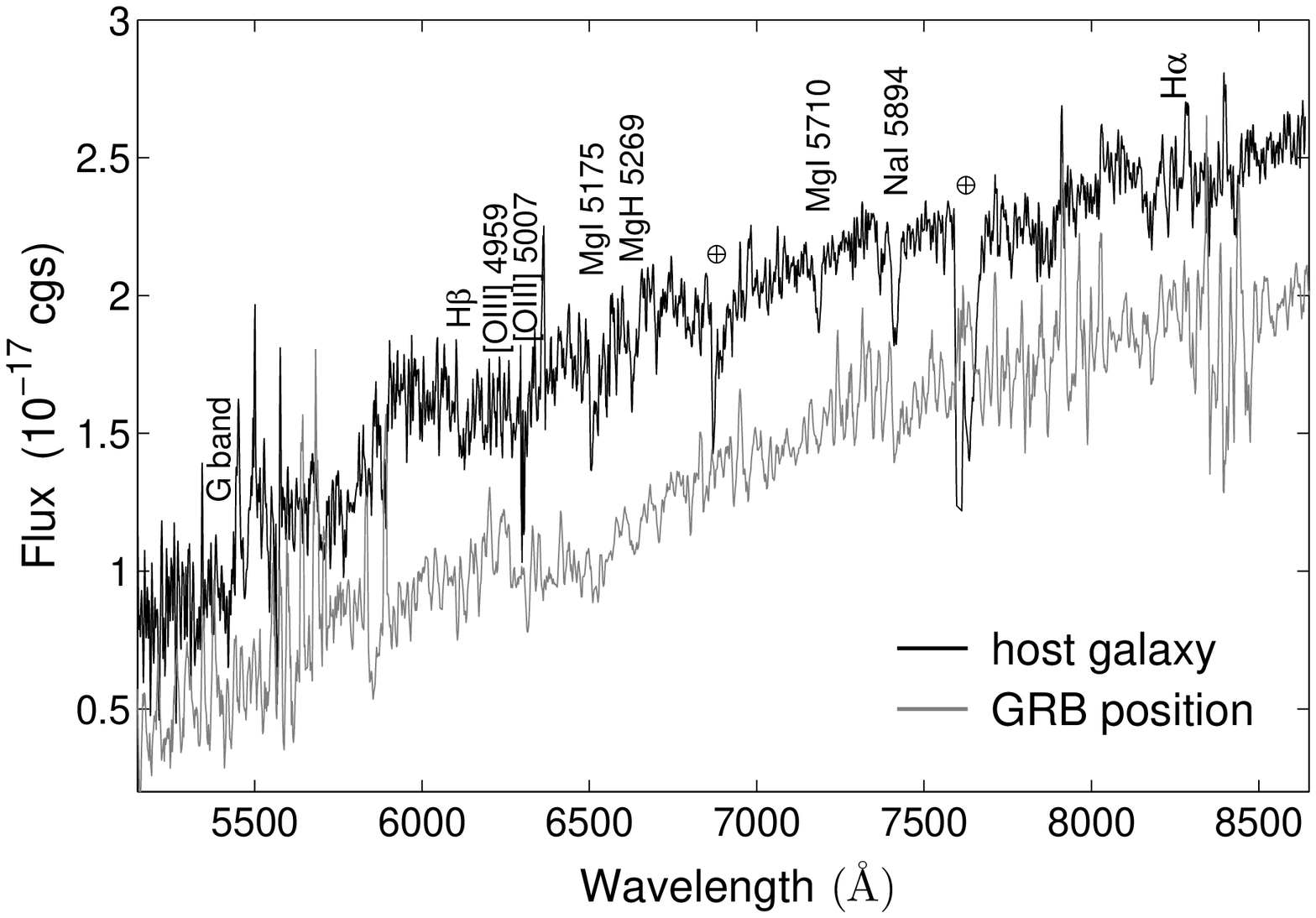}}
\resizebox{16pc}{!}{\includegraphics[height=.5\textheight]{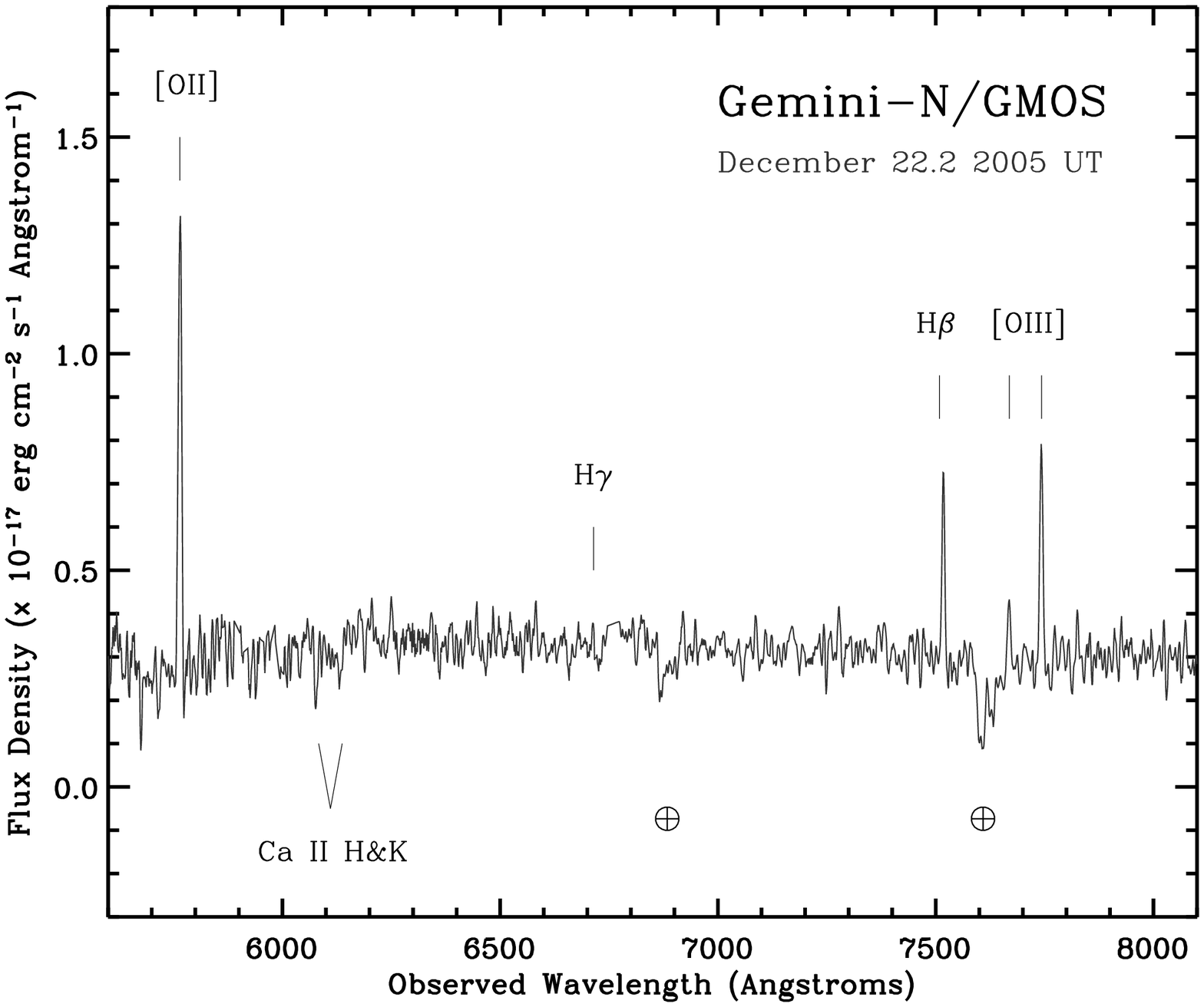}}
\caption{Spectra of the host galaxies of GRB\,050724 (left) exhibiting 
typical features of an early-type galaxy, and GRB\,051221 (right)
exhibiting emission lines typical of star forming galaxies.  From
Refs.~\cite{berger:05,soderberg:06}.}
\label{fig:spec}
\end{figure}

It has also been noted that some SHBs occur in galaxy clusters.
GRB\,050509b is likely located in the cluster ZwCl 1234.0+02916, while
GRB\,050813 is most likely associated with a high redshift cluster.
Several other cluster associations have been claimed for SHBs that
lack arcsecond positions.  At the present it is difficult to assess
whether the latter associations are in fact correct, but some caution
should be exercised given that the probability of finding a cluster in
the NED archive within $10'$ from a random position on the sky is
non-negligible, $\sim 5\%$ \cite{gcn:4553}.  Moreover, since none of
the SHBs with arcsecond positions are located in clusters, we consider
the cluster associations claimed to date to be suggestive, though not
secure.

Finally, as in the case of long GRBs the offset distribution has been
proposed as a test of the progenitor population.  The observed offsets
in the coalescence model are a function of progenitor lifetimes,
velocity kicks, and galaxy masses.  Predictions from population
synthesis models vary considerably, with possibly $50\%$ of the bursts
predicted to occur at offsets of $>10$ kpc \cite{fryer:99,perna:02}.
Observationally, the offsets appear to be relatively small, $\simlt
{\rm few}$ kpc \cite{berger:05,fox:05,soderberg:06}, but we stress
that given the uncertain input distributions, this test may not be as
useful as originally envisioned.

\subsection{Event Rates and Progenitor Lifetimes}

Another set of useful constraints on SHB models may be obtained from
the event rates and the lifetime of the progenitors.  Several authors
have attempted to constrain the progenitor lifetime distribution using
the BATSE flux distribution and the observed redshifts, leading to
conflicting results.  The canonical time delay distribution, $P(\tau)
\propto 1/\tau$ as inferred from Galactic DNS systems, may be
disfavored by the relatively low redshifts of some SHBs
\cite{gal-yam:05,nakar:05}, but is not ruled out \cite{guetta:05}.  
Other lifetime distributions (with a constant delay or constant rate)
also provide adequate fits.  The redshift distribution, particularly
when augmented by claimed associations for IPN SHBs \cite{gal-yam:05},
has been used to argue for lifetimes of $\sim 6$ Gyr with a relatively
small spread \cite{nakar:05}.  We caution here that the IPN
associations may be spurious\footnote{The ``brightest galaxy''
criterion clearly fails for some of the precisely localized SHBs.} and
in particular bias the result to longer lifetimes.  Moreover, with
$z\approx 1.8$ for GRB\,050813 the progenitor lifetime is constrained
to be $\simlt 3$ Gyr.

A more profitable approach may be to investigate the relative fraction
of SHBs in early- and late-type galaxies.  For SNe Ia, which are more
prevalent in late-type galaxies, similar analyses have led to the
conclusion that the progenitor lifetimes are $\tau\sim 1-3$ Gyr
\cite{tonry:03}.  In the case of SHBs, several authors have argued that 
the fraction in early-type galaxies is larger, and therefore the
progenitors lifetimes are longer than for SNe Ia, i.e. $\simgt 3$ Gyr
\cite{gal-yam:05}.  This analysis is highly uncertain at the present,
however, since of the three precisely localized SHBs, two are in fact
located in late-type galaxies.

Finally, the SHB local event rate is inferred to be at least $R\sim
10$ Gpc$^{-3}$ yr$^{-1}$ based on the BATSE rate and the current
redshift distribution \cite{nakar:05,guetta:05}.  The true rate is
likely higher due to beaming, introducing a correction of
$f_b^{-1}\sim 10-100$ for typical angles of $8-25^\circ$.  Another
unknown correction is due to the low-luminosity cutoff, $R\propto
L_{\rm min}^{-1}$.

\section{Summary and Future Directions}

The progress in our understanding of SHBs over the last several months
has shifted the field from the realm of speculation to a quantitative
study.  The observations made to date have allowed us to determine
some of the most important and basic properties:
\begin{itemize}
\item SHBs occur at cosmological distances with a relatively wide 
spread in redshift; a local population may contribute up to $20\%$ of
the BATSE sample.
\item The energy scale is typically lower than that of long GRBs, 
ranging from ${\rm few}\times 10^{48}$ erg to $\simgt 10^{50}$ erg.
\item The ejecta appear to be collimated, but with opening angles 
that are larger than the median for long GRBs.
\item SHBs tend to occur in lower density environments than long 
GRBs.
\item SHBs occur in elliptical and star forming galaxies, with 
roughly equal numbers.
\end{itemize}

These properties are in broad agreement with the DNS or NS-BH
coalescence models, but other possibilities remain equally viable.  In
particular the observed long-term flares and injection episodes, as
well as the spread in energies and host types, may point to multiple
progenitor systems and/or energy extraction mechanisms.  Clearly, an
increased sample of events will shed additional light on the
progenitor population.  Future tests may include studies of the
correlation between burst and host properties, evolution of the burst
properties as a function of redshift, and perhaps the detection of
gravitational waves in coincidence with an SHB.  An intriguing
possibility if SHBs in fact tend to be associated with galaxy clusters
is that they may provide a beacon to select clusters at high redshift,
$z\simgt 1.5$, where blind searches are exceedingly difficult.  This
may already be the case with GRB\,050813.


\begin{theacknowledgments}
It is a pleasure to thank my collaborators for their hard work in
pursuit of an understanding of short GRBs, in particular Derek Fox,
Dale Frail, Mike Gladders, Shri Kulkarni, and Alicia Soderberg.  This
work was supported by a Hubble Post-doctoral Fellowship grant,
HST-HF-01171.01.

\end{theacknowledgments}

\hyphenation{Post-Script Sprin-ger}


\end{document}